\documentclass[aps,onecolumn,showpacs]{revtex4}
\usepackage{amsmath,graphicx,subfig}
\usepackage{caption}
\usepackage{amssymb,amsbsy}

\begin{document}

\begin{center}{\large{\bf First Order Actions for New Massive Dual Gravities }}
\vskip 0.5cm
Alexangel Bracho and Adel Khoudeir
\\ \vskip 0.2cm
{\it Centro de F\'{\i}sica Fundamental, Departamento de F\'{\i}sica, \\
Facultad de Ciencias, Universidad de Los Andes, M\'erida 5101, Venezuela}
\vskip 0.3cm
{\it e-mail: alexangelb@hotmail.com, adel@ula.ve}
\end{center}
\vskip 1cm
\begin{abstract}
We present a first order formulation for the fourth order action of the new massive dual gravity in four dimensions. This proposal is easily generalized to arbitrary dimension. Also, we obtain the dual actions for massless and massive Curtright fields in D dimensions.
\end{abstract}



\section{INTRODUCTION}

Higher order derivative theories usually contain ghost excitations,
which make them non - unitary \cite{pu}, although other quantum properties can be improved \cite{thirring}. This is the
case for gravities with curvature squared terms \cite{stelle}.
There are exceptions. For instance, some scalar fields coupled to gravitation admit a higher order action description
($R + R^2 $ theory, see \cite{schmidt}). Also, the action for ``Galileons" \cite{nrt} contains higher derivatives, however their field equations are non-linear second order differential equations. This is analogous to the Lovelock gravity \cite{lovelock}, whose second order field equations
arise from an action which is higher order in curvature.
In three dimensions, it is possible to have higher order gravity theories without loss of unitarity. The non dynamical Einstein-Hilbert action can be augmented with a third order derivative Lorentz-Chern-Simons term, the Topologically Massive Gravity theory\cite{djt}, which describes the local propagation of one, ghost free and parity sensitive excitation: a massive graviton. Furthermore, the dubbed New Massive Gravity in three dimensions
was formulated by Bergshoeff, Hohm and Townsend \cite{bht} four years ago. It consists, again, in the non-dynamical Einstein-Hilbert action supplemented with a specific curvature squared term, which leads to fourth order field equation. At linearized level, the New Massive Gravity is equivalent to the standard massive spin 2 Fierz-Pauli theory, which propagates two degrees of freedom with the same mass and opposite helicities $\pm 2$. In both cases, the Einstein-Hilbert action has a wrong sign, which is essential in order to have free ghost excitations. The New Massive Gravity has a discontinuity in its degrees of freedom when the massless limit is considered. This limit leads to the fourth order Schouten gravity, which describes a massless, conformal invariant, ghost free local excitation \cite{deser}.
Also, the General Massive Gravity \cite{abrhst}, the most general gravity theory in three dimensions, propagates two massive gravitons of helicities $\pm 2$ but with different masses. Its gauge invariant linearized action was shown to be dual equivalent \cite{aks} to the massive spin 2 Fierz-Pauli with a term which breaks explicitly the local Lorentz symmetry \cite{aak1}.

Remarkably enough, it is the observation made in \cite{bht3} and \cite{bkrty} that solving on shell algebraic and differential constraints, it is possible to obtain in three dimensions higher order field equations for higher spin. This procedure was extended to four dimensions, but considering the massive Curtright field \cite{curtright}, which is the dual field to the massive spin 2 Fierz-Pauli theory in four dimensions
\cite{zinoviev}, \cite{gkmu}.
In consequence, a fourth order derivative action for the massive Curtright field is obtained.
A necessary condition is that the field, which solves the constraints, belongs to the same Lorentz representation as the original field.
This theory was called New Massive Dual Gravity \cite{bf-mrt}. This is an unitary theory by construction. The extension to arbitrary dimension has been achieved recently \cite{jmk}, \cite{ds}.

The aim of this work is to present a first order formulation for New Massive Dual Gravities.
The plan of this work is the following: in the next section, we introduce a first order formulation for a massless, mixed symmetry $\Phi_{mn,p}$ field, valid for arbitrary dimensions. From this action, we obtain the corresponding dual theory. In section 3, we perform the dimensional reduction of the action presented in section 2, keeping only first massive modes, to obtain the first order formulation for the massive $\Phi_{mn,p}$ field. In section 4, we propose the first order action for the New Massive Dual Gravity in four dimensions. The generalization to any dimension D  will be straightforward. Throughout this work we use $\eta_{mn}$ mostly positive and brackets indicate anti-symmetrization without any normalization factor, e. g. $V_{[mnp]} = V_{mnp} + V_{npm} + V_{pmn}$ .

\section{THE CURTRIGHT FIELD}
For general mixed symmetry fields $\Phi_{M_1 M_2 ... M_i ,N}$, with second order actions, Zinoviev has achieved first order actions \cite{zinoviev2}.
For our purpose, we will start with the following first order action for the massless field $\Phi^{MN,P} = -\Phi^{NM,P}$ in D dimensions
\begin{equation}\label{1}
 I = \int d^D x (\frac{3}{4}Y_{MNP,Q}Y^{MNQ,P} - \frac{3}{4}\frac{1}{(D-3)}Y^{MN}Y_{MN} - \frac{1}{2}Y^{MNP,Q}F_{MNP,Q}),
\end{equation}
where $Y^{MNP,Q}$ is an auxiliary field ($Y^{MN} = Y^{MNP},_P$) and $F_{MNP,Q}$ is the field strength of the field $\Phi^{MN,P}$
\begin{equation}
 F_{MNP,Q} = \partial_M \Phi_{NP,Q} + \partial_N \Phi_{PM,Q} + \partial_P \Phi_{MN,Q}.
\end{equation}
This action differs from the Zinoviev action by a change of variables:
\begin{equation}
Y_{MNP,Q} = \Omega_{Q,MNP} - \eta_{[QM}\Omega_{NP]}
\end{equation}
We have invariance under the following gauge transformations
\begin{equation}
\delta \Phi_{MN,P} = \partial_M z_{NP} - \partial_N z_{MP},
\end{equation}
being the gauge parameters $z_{MN}$ an arbitrary general second order tensor.
Besides, we have invariance under local "Lorentz" transformations
\begin{equation}
\delta \Phi_{MN,P} = \Lambda_{MNP}
\end{equation}
and
\begin{equation}
\delta Y_{MNP,Q} = \frac{1}{3}\partial_Q \Lambda_{[MNP]} - \partial_R \eta_{Q[M}\Lambda_{NP]R}
\end{equation}
where the parameters $\Lambda_{MNP}$ are completely antisymmetric.

We decompose the $\Phi^{MN,P}$ field as
\begin{equation}
\Phi^{MN,P} = T^{MN,P} + C^{MN,P},
\end{equation}
where $T^{MN,P} = -T^{NM,P}$ is the Curtright field, which satisfies the cyclic identity
$T_{[MN,P]} \equiv 0$ and $C_{MNP}$ is completely antisymmetric. The Lorentz symmetry indicates that the
$C_{MNP}$ is a non-dynamical field and the action will be written down only in terms of the Curtright field. Let us see this.
The equation of motion obtained after independent variations of the auxiliary field $Y^{MNP,Q}$ is
\begin{equation}
Y_{Q[MN,P]} - \frac{1}{D-3} \eta_{Q[M} Y_{NP]} = F_{MNP,Q}.
\end{equation}
This equation can be solved for $Y_{MNP,Q}$ as
\begin{equation}
Y_{MNP,Q} = \frac{2}{3}F_{MNP,Q} + \frac{1}{3} F_{Q[MN,P]} - \eta_{Q[M}F_{NP]}.
\end{equation}
Substituting back into (\ref{1}) we obtain the second order action
\begin{equation}
 I = \int d^D X[ -\frac{1}{6}F_{MNP,Q}F^{MNP,Q} -\frac{1}{4}F_{MNP,Q}F^{MNQ,P} + \frac{3}{4}F^{MN}F_{MN} ],
\end{equation}
where the field strength is now expressed only in terms of the Curtright field $T_{MN,P}$.
Equivalently, after using the property of cyclic identity satisfied by $T_{MN,P}$ and omitting divergences, this last
action can be rewritten as originally proposed by Curtright:
\begin{equation}
 I = \frac{3}{2} \int d^D X[ -\frac{1}{6}F_{MNP,Q}F^{MNP,Q} + \frac{1}{2}F^{MN}F_{MN} ],
\end{equation}
On the other hand, we can consider the field equation obtained making independent variations on
$\Phi_{MN,P}$
\begin{equation}
\partial_Q Y^{QMN,P} = 0.
\end{equation}
Locally this equation can be solved in terms of a $\Phi_{S_1 ...S_{D-4}, Q}$ field as
\begin{equation}
Y_{PMN,Q} = \frac{1}{2} \epsilon^{PMNRS_1 ...S_{D-4}} \partial_R \Phi_{S_1 ...S_{D-4}, Q} \equiv \frac{1}{2(D-3)} \epsilon^{PMNS_1 ...S_{D-3}}F_{S_1 ...S_{D-3},Q}.
\end{equation}
We have defined $F_{S_1 ...S_{D-3},Q} \equiv \partial_{S_1} \Phi_{S_2 ...S_{D-3}, Q} +$ cyclic permutation.
In five dimensions, we have $Y_{PMN,Q} \sim \epsilon^{PMNRS} \partial_R \Phi_{S,Q}$ and plugging into (\ref{1}), the
linearized Einstein action is obtained, illustrating the well known duality relation between Curtright and the
massless spin 2 fields in five dimensions \cite{bch}. In six dimensions the Curtright field is self dual  and in general D dimensions, we have the following duality relationship
\begin{equation}
T_{MN,P} \Leftrightarrow \Phi_{M_1 ...M_{D-4}, N}.
\end{equation}
The dual action is
\begin{eqnarray}
I = \int d^D x [&-& \frac{(D-4)}{2(D-3)}F^{M_1 ...M_{D-3},N} F_{M_1 ...M_{D-3},N} - \frac{1}{2}F_{M_1 ...M_{D-4}N,P}F^{M_1 ...M_{D-4}P,N} \nonumber \\
&+& \frac{1}{2(D-3)}F_{M_1 ...M_{D-4}N,N}F_{M_1 ...M_{D-4}P,P} .
\end{eqnarray}

\section{DIMENSIONAL REDUCTION}

In this section we perform a dimensional reduction to the action (\ref{1}) from D to D-1 dimensions in order to provide mass to the
Curtright field. We will keep only the first massive modes in a similar way as was accomplished for spin 2 in \cite{kmu}.
For this goal, we make the following definitions:
\begin{equation}
Y_{mnp,q(x,y)} = \sqrt{\frac{\mu}{\pi}} Y_{mnp,q(x)} \cos \mu y ,
\end{equation}
\begin{equation}
Y_{mny,q(x,y)} = \sqrt{\frac{\mu}{\pi}} Y_{mn,p(x)} \sin \mu y ,
\end{equation}
\begin{equation}
Y_{mnp,y(x,y)} = \sqrt{\frac{\mu}{\pi}} X_{mnp(x)} \sin \mu y ,
\end{equation}
\begin{equation}
Y_{mny,y(x,y)} = \sqrt{\frac{\mu}{\pi}} Z_{mn(x)} \cos \mu y
\end{equation}
and
\begin{equation}
\Phi_{mn,p(x,y)} = \sqrt{\frac{\mu}{\pi}} \Phi_{mn,p(x)} \cos \mu y ,
\end{equation}
\begin{equation}
\Phi_{my,n(x,y)} = \sqrt{\frac{\mu}{\pi}} \Phi_{mn(x)} \sin \mu y ,
\end{equation}
\begin{equation}
\Phi_{mn,y(x,y)} = \sqrt{\frac{\mu}{\pi}} B_{mn(x)} \sin \mu y ,
\end{equation}
\begin{equation}
\Phi_{my,y(x,y)} = \sqrt{\frac{\mu}{\pi}} a_{m(x)} \cos \mu y .
\end{equation}
The dependence with the extra compact dimension is denoted by $y$, and low letters: $m,n,p,...$ indicate
coordinates in the reduced spacetime. The fields $X_{mnp}, Z_{mn}$, $\Phi_{mn}$ and $B_{mn}$ are completely antisymmetric, $a_m $ is a vector
and $Y_{mnp,q}$  and $Y_{mn,p}$ will play the role of auxiliary fields.
Substituting these definitions and performing the integration on the compact coordinate we obtain the dimensionally reduced action
$y$. The result is
\begin{eqnarray}
 I &=& \int d^{D-1} x [\frac{3}{4}Y_{mnp,q}Y^{mnq,p} - \frac{3}{4(D-3)}Y_{mnp,p}Y^{mnq,q} - \frac{3}{2}Y^{mnp,q}\partial_m \Phi_{np,q} \nonumber \\
  &-& \frac{3}{2(D-3)}Y^{mnp,p}Z_{mn} + \frac{3}{4}\frac{D-4}{D-3}Z^{mn}Z_{mn} - \frac{3}{2}Z^{mn}(\partial_m a_n -\partial_n a_m + \mu B_{mn}) \nonumber \\
  &-&\frac{3}{2}Y^{mn,p}(\partial_m \Phi_{np} - \partial_n \Phi_{mp} - \mu \Phi_{mn,p})
  - \frac{1}{2}X^{mnp} (\partial_m B_{np} + \partial_n B_{pm} + \partial_p B_{mn})] \nonumber \\
  &+&\frac{3}{2}Y^{mn,p}Y_{mp,n} - \frac{3}{2(D-3)}Y^{mp,p}Y_{mq,q} + \frac{3}{2}X^{mnp}Y_{mp,n} ].
\end{eqnarray}
This reduced action is invariant under the following gauge transformations
\begin{equation}
\delta \Phi_{mn,p(x)} = \partial_m z_{np(x)} - \partial_n z_{mp(x)},
\end{equation}
\begin{equation}
\delta B_{mn} = \partial_m z_{ny} - \partial_n z_{my},
\end{equation}
\begin{equation}
\delta \Phi_{mn} = \partial_m z_{yn} + \mu z_{mn}
\end{equation}
and
\begin{equation}
\delta a_ m = \partial_m z_{yy} - \mu  z_{my},
\end{equation}
where we split up the gauge parameters as: $z_{MN} : (z_{mn}, z_{my}, z_{ym}, z_{yy})$.

We can break these gauge symmetries choosing appropriately the $z_{mn}$ and $z_{my}$ gauge parameters to fix the gauges
\begin{equation}
\Phi_{mn} = 0  \quad and \quad a_m = 0.
\end{equation}
The $\Phi_{mn}$ and $a_m $ are Stueckelberg fields. Furthermore, the reduced action inherited "Lorentz" symmetries with parameters
$\Lambda_{MNP}: (\Lambda_{mnp}, \Lambda_{mny} \equiv \Lambda_{mn})$, expressed in the following transformations:
\begin{equation}
\delta \Phi_{mn,p} = \Lambda_{mnp},
\end{equation}
\begin{equation}
\delta \Phi_{mn} = -\Lambda_{mn} = -\delta B_{mn}
\end{equation}
\begin{equation}
\delta a_{m} = 0,
\end{equation}
\begin{equation}
\delta Y_{mnp,q} = \partial_q \Lambda_{mnp} - \partial_r \eta_{q[m}\Lambda_{np]r} - \mu \eta_{q[m}\Lambda_{np]},
\end{equation}
\begin{equation}
\delta Y_{mn,p} = \partial_p \Lambda_{mn} + \eta_{p[m}\partial_q \Lambda_{n]q},
\end{equation}
\begin{equation}
\delta X_{mnp} = - \mu \Lambda_{mnp}
\end{equation}
and
\begin{equation}
\delta Z_{mn} = - \partial_p \Lambda_{pmn}.
\end{equation}
Similarly, we can break the Lorentz symmetry associated with the $\Lambda_{mn}$ parameter by choosing
\begin{equation}
B_{mn} = 0
\end{equation}
At this stage, we can eliminate the $Z_{mn}$ field since it appears as a quadratic multiplier. Its equation of motion leads to
determining its value:
\begin{equation}
 Z_{mn} = \frac{1}{D-4}Y_{mnp,^{p}} .
\end{equation}
With these gauges fixing and substituting this value of $Z_{mn}$, the reduced action boils down to
\begin{eqnarray}\label{2}
 I &=& \int d^D x[\frac{3}{4}Y_{mnp,q}Y^{mnq,p} - \frac{3}{4(D-3)}Y_{mnp,p}Y^{mnq,q} \nonumber \\
 &-& \frac{3}{2}Y^{mnp,q}\partial_m \Phi_{np,q} + \frac{3}{2}X^{mnp} Y_{np,m} \nonumber \\
  &+&\frac{3}{2}Y^{mn,p}Y_{mp,n} - \frac{3}{2(D-2)}Y^{mp,p}Y_{mq,q}
  +\frac{3\mu}{2}Y^{mn,p}  \Phi_{mn,p}].
\end{eqnarray}
The local Lorentz symmetry associated with the parameter $\Lambda_{mnp}$ can be used to fix gauge in two different ways. The first alternative
consists in choosing $X^{mnp} = 0$ and the other possibility allows us to eliminate the antisymmetric part of
$\Phi_{mn,p}$ (i. e.  $\Phi_{[mn,p]} \equiv 0$). Both choices will lead to the
same result: the first order formulation for the massive Curtright action. Fixing the gauge
$X^{mnp} = 0$, the field equation for the $Y^{mn,p}$ is
\begin{equation}
Y_{mp,n} - Y_{np,m} - \frac{1}{D-2} (\eta_{pn} Y_{m} - \eta_{pm} Y_{n} )= -\mu \Phi_{mn,p},
\end{equation}
which can be solved as
\begin{equation}
Y_{mn,p} = -\frac{\mu}{2} [\Phi_{mn,p} + \Phi_{mp,n} - \Phi_{np,m}] - \mu[\eta_{mp}\Phi_n - \eta_{np}\Phi_m]
\end{equation}
and when this expression is plugged into (\ref{2}), the result is the following non-gauge invariant action
\begin{eqnarray}\label{3}
 I &=& \int d^D x [\frac{3}{4}Y_{mnp,q}Y^{mnq,p} - \frac{3}{4(D-3)}Y_{mnp,p}Y^{mnq,q} - \frac{3}{2}Y^{mnp,q}\partial_m \Phi_{np,q} \nonumber \\
  &-&\frac{3\mu^{2}}{8} (\Phi^{mn,p}\Phi_{mp,n} + 2\Phi^{mn,p}\Phi_{mp,n} -4\Phi^{m}  \Phi_{m}) ].
\end{eqnarray}
The first line in this dimensional reduced action (\ref{3}) is the first order action for the massless Curtright field.
Now, if we decompose the $\Phi_{mn,p}$ in its irreducible parts : $\Phi_{mn,p} = T_{mn,p} + C_{mnp}$,
with $T_{[mn,p]} \equiv 0$ and $C_{mnp}$ completely antisymmetric, this part does not depends on
$C_{mnp}$, while the coefficients of the massive term, the second line in action (\ref{3}), are such that the constituents of $\Phi_{mn,p}$
are decoupled. Indeed
\begin{equation}
\Phi^{mn,p}\Phi_{mn,p} + 2\Phi^{mn,p}\Phi_{mp,n} -4\Phi^{m}  \Phi_{m} = 2T_{mn,p}T^{mn,p} - 4T_m T^m - C_{mnp}C^{mnp}.
\end{equation}
In consequence, $C_{mnp}$ is a non dynamical field, and then by eliminating $Y_{mnp,q}$ through its equation of motion, we obtain
the second order action for the massive Curtright field.
\begin{equation}\label{massive}
 I = \frac{3}{2} \int d^D x[ -\frac{1}{6}F_{mnp,q}F^{mnp,q} + \frac{1}{2}F^{mn}F_{mn} - \frac{\mu^{2}}{2} (T_{mn,p}T^{mn,p} - 2T_m T^m )].
\end{equation}
The number of degrees of freedom of this theory is given by $\frac{1}{3}(D-1)(D+1)(D-3)$, while in the massless case the Curtright field
propagates $\frac{1}{3}D(D-2)(D-4)$ transverse modes. The massless Curtright field in D + 1 dimensions has the same number of degrees of
freedom as the massive Curtright field in D dimensions.

If we have chosen the gauge $C_{mnp} = 0$ $(\Phi_{mn,p} = T_{mn,p})$, then $X^{mnp}$ is a multiplier which tells us that
$Y_{[mn,p]} = 0$ After eliminating $Y_{mn,p}$ through the use of its equation of motion, we again reach the action for the massive
Curtright field. From now on, we will refer only to the Curtright $T_{mn,p}$.

From the first order action (\ref{3}), we can obtain the dual theory for the massive Curtright field if we eliminate the $\Phi^{mn,p}$ field through
its equation of motion. In fact, we obtain
\begin{equation}\label{29}
\Phi^{mn,p} = -\frac{1}{\mu^2 }\partial_q (Y^{qpm,n} - Y^{qpn,m}) + \frac{1}{\mu^2 (D-2)}\partial_q (\eta^{pm}Y^{qn} - \eta^{pn}Y^{qm})
\end{equation}
and substituting into (\ref{3}), the following action for $Y^{mnp,q}$ is obtained
\begin{eqnarray}\label{33}
 \frac{2}{3}\mu^2 I &=& \int d^D x [\partial_q Y^{qmn,p}\partial^r Y_{rmp,n} - \frac{1}{D-2}\partial_q Y^{qm}\partial_r Y^{rm} \nonumber \\
  &+&\frac{\mu^{2}}{2} (Y^{mnp,q}Y_{mnq,p} - \frac{1}{D-3}Y^{mn}Y_{mn} )].
\end{eqnarray}
Now, if we decompose $Y^{mnp,q}$ in its trace and traceless parts
\begin{equation}
Y^{mnp,q} = W^{mnp,q} + \frac{1}{D-2}(\eta^{qm}Y^{np} + \eta^{qn}Y^{pm} + \eta^{qp}Y^{mn}),
\end{equation}
where $W^{mnp,q}$ is the traceless part of $Y^{mnp,q}$ ($W^{mnp,} _p = 0$), the action is written down as
\begin{equation}
\frac{2}{3}\mu^2 I = \int d^D x [\partial_q W^{qmn,p}\partial^r W_{rmp,n}
  +\frac{\mu^{2}}{2} (W^{mnp,q}W_{mnq,p} - \frac{1}{(D-2)(D-3)}Y^{mn}Y_{mn}) ],
\end{equation}
which clearly shows that the trace of $Y^{mnp,q}$ is a non dynamical variable. Next, we introduce the dual of $W^{mnp,q}$:
\begin{equation}
W^{mnp,q} = \frac{1}{(D-3)}\epsilon^{mnpr_1 ... r_{D-3}} T_{r_1 ... r_{D-3} q} .
\end{equation}

Since $W^{mnp,q}$ is traceless, then $T_{r_1 ... r_{D-3}, q}$ satisfy the cyclic identity: $T_{[r_1 ... r_{D-3} ,q]} \equiv 0$ and introducing the field strength $F_{r_1 r_2 ...r_{D-2}, q} \equiv \partial_{r_1} T_{r_2 ... r_{D-3} ,q} +$ cyclic permutations, the dual action for the massive Curtright field is
\begin{eqnarray}
I = \int d^D x [&-& \frac{1}{(D-2)}F^{r_1 ...r_{D-3}p,q} F_{r_1 ...r_{D-3}p,q} - F_{r_1 ...r_{D-3}p,p}F^{r_1 ...r_{D-4}q,q} \nonumber \\
&-& \mu^2 (T_{r_1 ...r_{D-3},p}T^{r_1 ...r_{D-3},p} - (D-3)T_{r_1 ...r_{D-4}p,p}T^{r_1 ...r_{D-4}p,p}.
\end{eqnarray}

Thus, we have established the duality relationship between the massive Curtright field ($T_{mn,p}$) and a massive mixed symmetry field
$T_{m_1 ... m_{D-3},p}$

\section{THE FIRST ORDER ACTION}
In four dimensions the massless Curtright does not propagate any local excitations like the massless spin 2 in three dimensions. Generally, a massless $\Phi_{m_1 ,m_2, m_{D-2}, n}$ field in D dimensions does not have any degrees of freedom. In four dimensions,
Zinoviev has written down a first order action for the non dynamical Curtright field, akin to the well known first order action of linearized Einstein action in three dimensions. This action is expressed in terms of the $T_{mn,p}$  field and an auxiliary field $h_{mn}$ neither symmetric nor antisymmetric. This action is written out as
 \begin{equation}\label{69}
 I_C = \int d^4 x [-\frac{1}{4}(h_{mn}h^{nm} - h^2 ) + \frac{1}{2}\epsilon^{mnpq}h_{mr}\partial_n T_{pq,r} ].
 \end{equation}
 In fact, the field equation obtained by varying the auxiliary field $h^{nm} $ tells us:
\begin{equation}
h^{nm} =  \epsilon^{mpqr}\partial_p T_{qr,n} .
\end{equation}

Note that $h = 0$ on shell, but we will keep the $h^2$ term in the action because it will be necessary  to have the dual off-shell equivalence between massive spin 2 and Curtright fields in four dimensions. When this value of $h_{nm}$ is substituted into $I_C$, the second order Curtright action is obtained
\begin{equation}
I_C =  \frac{1}{2} \int d^4 x T_{mn,p} G^{mn,p},
\end{equation}
where
\begin{equation}\label{5}
G^{mn,p} \equiv \frac{1}{2}\epsilon^{mnsr}\epsilon^{pquv}\partial_r \partial_q T_{uv,s}
\end{equation}
is the "generalized Einstein tensor" for the Curtright field introduced in ref \cite{bf-mrt}.

Now, we are ready to propose the first order action in four dimensions for the fourth order "new massive gravity".
This action involves three independent variables: the Curtright field $T_{mn,p}$, a $h_{mn}$ field and an auxiliary field $Y_{mn,p}$. Both $T_{mn,p}$ and $Y_{mn,p}$ satisfy cyclic identities. The action is written down as:
\begin{eqnarray}\label{96}
 I = \int d^4 x [&-&\frac{1}{4}(h_{mn}h^{nm} - h^2 ) + \frac{1}{2}\epsilon^{mnpq}h_{mr}\partial_n T_{pq,r} \nonumber \\
 &+& Y^{mn,p}Y_{mp,n} - \frac{1}{2} Y^m Y_m + \mu Y^{mn,p}T_{mn,p} ].
 \end{eqnarray}
Making independent variations on the fields, we obtain
\begin{equation}\label{a}
\frac{\delta I}{\delta h_{nm}} = 0  \quad \Rightarrow  \quad h^{mn}  = \epsilon^{npqr}\partial_p T_{qr,m},
\end{equation}
\begin{equation}\label{b}
\frac{\delta I}{\delta Y_{mn,p}} = 0  \quad \Rightarrow  \quad Y_{mn,p} = -\frac{\mu}{2} [T_{mn,p} + T_{mp,n} - T_{np,m}] - \mu[\eta_{mp}T_n - \eta_{np}T_m]
\end{equation}
and
\begin{equation}\label{c}
\frac{\delta I}{\delta T_{mn,p}} = 0  \quad \Rightarrow  \quad Y^{mn,p} = -\frac{1}{2\mu}\epsilon^{mnqr}\partial_q h_{rp}.
\end{equation}
The equations (\ref{a}) and (\ref{c}) are first order in derivatives.
As usual, the auxiliary fields ($h_{mn}$ and $Y_{mn,p}$) can be substituted, using equations (\ref{a}) and (\ref{b}), in order to
reach the second order action (\ref{massive}) for the massive Curtright field $T_{mn,p}$.

Alternatively, we can express the $Y_{mn,p}$ field (eq. (\ref{c})) in terms of the second derivative of the Curtright field $T_{mn,p}$ using eq. (\ref{a}):
\begin{equation}\label{Y}
Y_{mn,p} = -\frac{1}{\mu} G_{mn,p[T]}.
\end{equation}
Since $Y_{mn,p}$ satisfies a cyclic identity, we can write
\begin{equation}
Y^{mn,p}Y_{mp,n} - \frac{1}{2} Y^m Y_m = \frac{1}{2}Y^{mn,p}Y_{mn,p} - \frac{1}{2} Y^m Y_m .
\end{equation}

Plugging into the first order action (\ref{96}), the values of $h_{mn}$ and $Y_{mn,p}$ given by equations (\ref{a}) and (\ref{Y}),
we obtain the fourth order action of the new massive gravity in four dimensions :
\begin{equation}\label{bmrt}
I =  \frac{1}{2} \int d^4 x [- T_{mn,p} G^{mn,p} + \frac{1}{\mu^2} G^{mn,p}S_{mn,p}],
\end{equation}
where
\begin{equation}\label{12}
S_{mn,p} \equiv  G_{mn,p} - \frac{1}{2}(\eta_{np}G_m - \eta_{mp}G_n )
\end{equation}
is a generalized "Schouten" tensor.
Note the the wrong sign of the kinematical Curtright term, it has been flipped over. In this way we have implemented a first order parent action which establishes the equivalence between the second order massive Curtright theory and the higher order action of \cite{bf-mrt}.

Alternatively, we can arrive to the fourth order action from  a second order action. After integrating out the  $h_{mn}$ auxiliary field, we introduce a new  $v_{mn,p}$  auxiliary field, in order to rewrite the second order action for the massive Curtright field in the following way:
\begin{eqnarray}\label{99}
 I = \int d^4 x [&-&  \frac{1}{2}  v_{mn,p}\epsilon^{mnsr}\epsilon^{pquv}\partial_r \partial_q v_{uv,s}  
+ T_{mn,p} \epsilon^{mnsr}\epsilon^{pquv}\partial_r \partial_q v_{uv,s}  
\nonumber \\
 &+& Y^{mn,p}Y_{mp,n} - \frac{1}{2} Y^m Y_m + \mu Y^{mn,p}T_{mn,p} ].
 \end{eqnarray}
Now, the $T_{mn,p}$ field is a linear multiplier, whose constraint is solved by $Y_{mn,p} = -\frac{1}{\mu} G_{mn,p[v]}$ which lead to the fourth order action for $v_{mn,p}$.

Moreover, we can establish the dual equivalence between massive Curtright field and the massive spin 2 field, in four dimensions, from our parent first order action. Indeed, using only (\ref{c}), we obtain:
\begin{equation}
I = \int d^4 x [-\frac{1}{16}(C_{mn,p}C^{mn,p} -\frac{1}{16}C_{mn,p}C^{mp,n} + \frac{1}{4}C_{m}C^{m} )
- \frac{1}{4}(h_{mn}h^{nm} - h^2 ) ],
\end{equation}
where $C_{mn,p} \equiv \partial_m h_{np} - \partial_n h_{mp}$ and $C_m = C_{mn,^n}$. This is just the massive second order spin 2 Fierz-Pauli action.

The generalization to D dimensions is
\begin{eqnarray}\label{30}
I &=& \int d^D x [-\frac{1}{4}(h_{mn}h^{nm} - h^2 ) + \frac{1}{2}\epsilon^{mn s_1 ... s_{D-2}}\omega_{mr}\partial_n T_{s_1 ... s_{D-2},r} \nonumber \\
&+& Y^{m_1 ... m_{D-3} n,p}Y_{m_1 ... m_{D-3}  n,p} - \frac{1}{D-2} Y^{m_1 ... m_{D-3}} Y_{m_1 ... m_{D-3}} \nonumber \\
&+& \mu Y^{m_1 ... m_{D-2},p}T_{m_1 ... m_{D-2},p} ],
\end{eqnarray}
where $T_{m_1 ,m_2, ..., m_{D-2}, n}$ is the dual massive field of the massive $h_{mn}$ field \cite{gkmu}. $Y_{m_1 ,m_2, ..., m_{D-2}, n}$ is an auxiliary field ( $Y_{m_1 ,m_2, ..., m_{D-3}} = Y_{m_1 ,m_2, ..., m_{D-3}n, n}$). Both $T_{m_1 ,m_2, m_{D-2}, n}$ and $Y_{m_1 ,m_2, m_{D-2}, n}$ satisfy cyclic identities.
Now, we have the following equations
\begin{equation}\label{31}
h^{mn} = \epsilon^{npr_1 ... r_{D-2}} \partial_p T_{r_1 ... r_{D-2}, m}
\end{equation}
and
\begin{equation}\label{32}
Y^{m_1  ... m_{D-2}, n} = -\frac{1}{\mu} G^{m_1  ... m_{D-2}, n},
\end{equation}
where
\begin{equation}\label{33}
G^{m_1  ... m_{D-2}, n} = \frac{1}{2}\epsilon^{m_1  ... m_{D-2}qp}\epsilon^{nsr_1  ... r_{D-2}}\partial_q \partial_s T_{r_1 ... r_{D-2}, p},
\end{equation}
is the generalized Einstein tensor. Plugging (\ref{31}) and (\ref{32}) into (\ref{30}), the fourth order action (\cite{jmk}, \cite{ds}) is obtained \begin{equation}\label{34}
I = \int d^D x [\frac{1}{2}T_{m_1  ... m_{D-2}, n} G^{m_1 ... m_{D-2}, n} + \frac{1}{\mu^2 }G^{m_1 ... m_{D-2}, n}S_{m_1 ... m_{D-2}, n}  ],
\end{equation}
where
\begin{equation}\label{35}
S_{m_1 ... m_{D-2}, n} = G_{m_1 ... m_{D-2}, n} - \frac{1}{D-2}\eta_{n [m_{D-2}} G_{m_1 ... m_{D-3}]p,p},
\end{equation}
is the generalized Schouten tensor. Also, the first order action (\ref{30}) permits to show the dual equivalence
\begin{equation}
h_{mn} \Leftrightarrow T_{m_1  ... m_{D-2}, n},
\end{equation}
which was established in (\cite{gkmu}) for the massive linearized gravitation in arbitrary dimensions.

Finally, we consider the case for three dimensions. We must expect to recover the fourth order new massive gravity \cite{bht}.
The first order action to consider is
\begin{eqnarray}\label{97}
 I = \int d^3 x [&-&\frac{1}{2}(\omega_{mn}\omega^{nm} - \omega^2 ) + \epsilon^{mnpq}\omega_{mr}\partial_n e_{pq} \nonumber \\
 &+& \frac{1}{2}Y^{mn}Y_{nm} - \frac{1}{4} Y^2  + \mu Y^{mn}e_{mn} ].
 \end{eqnarray}
We have denoted the auxiliary fields by $\omega_{mn}$ (for convenience) and $Y_{nm}$, which are general second order tensor.
The coefficients have been slightly changed in concordance with the result of ref \cite{kmu}.
The following equations of motion are obtained
\begin{equation}\label{d}
\frac{\delta I}{\delta \omega_{nm}} = 0  \quad \Rightarrow  \quad \omega_{mn} - \eta_{nm}\omega + \epsilon_n  ^{pq}\partial_p e_{qn} = 0,
\end{equation}
\begin{equation}\label{e}
\frac{\delta I}{\delta Y_{mn}} = 0  \quad \Rightarrow  \quad Y_{mn} - \frac{1}{2}\eta_{mn}Y + \mu e = 0
\end{equation}
and
\begin{equation}\label{f}
\frac{\delta I}{\delta e_{mn}} = 0  \quad \Rightarrow  \quad Y_{mn,p} = -\frac{1}{\mu}\epsilon_{m}^{pq}\partial_p \omega_{qn}.
\end{equation}
The equation (\ref{d}) is solved as:
\begin{eqnarray}\label{x}
\omega_{mn} &=& \epsilon_n  ^{pq}\partial_p e_{qm} - \frac{1}{2}\eta_{mn}\epsilon^{pqr}\partial_p e_{qr} \nonumber \\
&\equiv& W_{mn[e]}
\end{eqnarray}
while the equation (\ref{e}) determines $Y_{mn}$
\begin{equation}\label{xx}
Y_{mn} = -\mu (e_{mn} - \eta_{mn}e).
\end{equation}
Inserting equations (\ref{x}) and (\ref{xx}) into the equation (\ref{97}), the  following action is obtained
\begin{equation}\label{101}
I = \int d^3 x [\frac{1}{2}e_{mq}\epsilon^{mnp}\partial_n W_{pq[e]} - \frac{\mu^2 }{2}(e_{mn}e^{nm} - e^2 ),
\end{equation}
and after introducing the symmetric and antisymmetric parts of $e_{mn}$ ($e_{mn} =  h_{mn} + \epsilon_{mnp}v^q $), we recognize (\ref{101}) as the action for the massive spin 2. The first term is the linearized Einstein action ($e_{mq}\epsilon^{mnp}\partial_n W_{pq[e]} = h_{mn}G^{mn}$, $G^{mn} = R_{mn} - \frac{1}{2}\eta_{mn}R$ is the linearized Einstein tensor) expressed only in terms of the symmetric part $h_{mn}$ of $e_{mn}$. The antisymmetric component $v^q $ appears decoupled in the massive term. Alternatively, as in the formulation for the Curtright field, we can express $Y_{mn}$ as
\begin{equation}\label{102}
Y_{mn} = -\frac{1}{\mu}\epsilon_m ^{pq}\partial_p W_{qn} = -\frac{1}{\mu}G_{mn[h]}
\end{equation}
and substituting (\ref{x}) and (\ref{102}) into the action, the new massive gravity is reached:
\begin{equation}\label{103}
I_{BHT} = \frac{1}{2}\int d^3 x [ -h_{mn}G^{mn} + \frac{1}{\mu^2 }(R_{mn}R^{mn} - \frac{3}{8}R^2 ) ].
\end{equation}
Thus, we have implemented a first order formulation for the fourth order new massive gravity in three dimensions. On the other hand, the action expressed only in terms of $\omega_{mn}$ is the same form as for the $e_{mn}$. In three dimensions, the massive spin 2 field is self-dual.

\section{CONCLUSIONS}
In this paper we have established a first order action for the fourth order action of the New Massive Gravity in four dimensions. In order to achieve this goal, two auxiliary fields were introduced, besides the Curtright field.
One of these auxiliary fields comes from the dimentional reduction of the first order action for the Curtright field, 
while the other auxiliary field appears when we consider a first order kinematical term for a general mixed symmetry $T_{s_1 ... s_{D-2},r}$ \cite{zinoviev}:
\begin{equation}\label{top}
I = \int d^D x [-\frac{1}{4}(h_{mn}h^{nm} - h^2 ) + \frac{1}{2}\epsilon^{mn s_1 ... s_{D-2}}\omega_{mr}\partial_n T_{s_1 ... s_{D-2},r} ].
\end{equation}
This action is ''topological" in the sense that it does not propagate any local degrees of freedom. 
The extension to arbitrary dimensions was straightforward formulated, including $D=3$ new massive gravity.
Furthermore, we have established the dual actions for massless and massive Curtright field in any dimensions:
\begin{equation}
T_{mn,p} \Leftrightarrow T_{m_1 ...m_{D-4}, p} (\mu = 0) \quad T_{mn,p} \Leftrightarrow T_{m_1 ...m_{D-3}, p} (\mu \neq 0).
\end{equation}

\section*{Acknowledgements}
We would like to thank you Denis Dalmazi for useful comments. Also, we would like to thank you to Willians Barreto and Luis Moreno Villamediana for reading and revising the manuscript.

\end{document}